\documentclass[journal=jacsat,manuscript=article]{achemso}

\usepackage[T1]{fontenc} 
\usepackage{dcolumn} 
\usepackage{graphicx}

\author{Souren Adhikary and Sudipta Dutta}
\affiliation{Department of Physics, Indian Institute of Science Education and Research (IISER) Tirupati, Tirupati - 517507, Andhra Pradesh, India}
\email{sdutta@iisertirupati.ac.in, sourenadhikary@students.iisertirupati.ac.in}

\title[An \textsf{achemso} demo]
  {Circular dichroism in two-dimensional BC$_6$N and B$_3$C$_2$N$_3$ in absence of intervalley excitonic coupling}

\begin{document}



\begin{abstract}
Two-dimensional (2D) noncentrosymmetric systems offer potential opportunities for exploiting the valley degrees of freedom for advanced information processing, owing to non-zero Berry curvature. However, such valley polarization in 2D materials is crucially governed by the intervalley excitonic scattering in momentum space due to reduced electronic degrees of freedom and consequent enhanced electronic correlation. Here, we study the valley excitonic properties of two 2D noncentrosymmetric complementary structures, namely, BC$_6$N and B$_3$C$_2$N$_3$ using first principles-based GW calculations combined with the Bethe-Salpeter equation (BSE), that brings the many-body interactions among the quasiparticles. The \textbf{k}-resolved oscillator strength of their first bright exciton indicates their ability to exhibit valley polarization under the irradiation of circularly polarized light of different chiralities. Both the systems show significant singlet excitonic binding energies of 0.74 eV and 1.31 eV, respectively. Higher stability of dark triplet excitons as compared to the singlet one can lead to higher quantum efficiency in both the systems. The combination of large excitonic binding energies and the valley polarization ability with minimal intervalley scattering make them promising candidates for applications in advanced optical devices and information storage technologies.
\end{abstract}

%
%
%
%
%

\section{Introduction:}

\noindent The two-fold valley degeneracy in the reciprocal space of a hexagonal two-dimensional (2D) semiconductor is an essential factor for the exploration of valleytronics\cite{rev-1,rev-2,rev-3,rev-4}. Uncovering the valley degeneracy between those two valleys could be utilized as `zeros' and `ones' in next-generation information storage devices\cite{binary-valley}. Noncentrosymmetric 2D systems are of great interest owing to their capability of valley polarization\cite{graphene-valley}. However, in these materials, the valley properties are predominantly influenced by excitons and their scattering processes, which involve the separation between valleys in momentum space\cite{hbn-prl-intervalley}.
In this context, monolayer semiconducting transition-metal dichalcogenides (TMDs) are prominent materials due to their broken spatial inversion symmetry and low-energy excitations are highly localized in valleys in the first Brillouin zone (BZ)\cite{TMD-1,TMD-2,TMD-3}. Apart from the d-block TMDs, valley polarization in p-block monolayers has been mostly elusive due to their lack of broken inversion symmetry\cite{graphene-valley}.

Graphene, the most extensively studied 2D system, does not possess any gaps at the valleys and maintains the spatial inversion symmetry that leads to zero Berry curvature, which limits its use in valley polarization\cite{neto-rev-modern-physics,geim2007rise}. Therefore, materials with gap will attract more attention. Monolayer hexagonal-boron nitride (h-BN) would be a perfect candidate due to its weak dielectric screening and large band gap, which result in a large binding energy of low-lying bound exciton\cite{hBN-valley-1}. In addition to the large binding energy of the first bright exciton, monolayer h-BN is also a noncentosymetric system that is expected to show valley polarization. However, its weak screening results in the coupling of free electron-hole pairs over a large area of the BZ. Due to this intervalley coupling, the exitons arising from different chirality of circular polarized lights at two valleys get mixed and fails to show valley polarization\cite{hbn-prl-intervalley}. The best materials for exhibiting valley polarization would be those with low-lying excitons that does not undergo any intervalley coupling.

Recently, there has been a growing interest in 2D borocarbonitrides, such as BC$_6$N and B$_3$C$_2$N$_3$, due to their remarkable properties for semiconductor device applications\cite{hBN-valley-1,bc6n-1,bc6n-2}. The BC$_6$N, for instance, is a highly promising material characterized by its high charge carrier mobility and a moderate direct band gap\cite{bc6n-1,karmakar2022strain}. Theoretical predictions suggest that BC$_6$N exhibits excellent gas sensing ability, photocatalytic capabilities, and shows promise as an anode material for potassium-ion batteries\cite{bc6n-1,karmakar2022strain,abdullah2021properties,rahimi2021hydrogen,xiang2020flexible}. On the other hand, monolayer B$_3$C$_2$N$_3$ also demonstrates potential for photo-catalytic abilities\cite{karmakar2023g}. Furthermore, BC$_6$N and B$_3$C$_2$N$_3$ possess broken spatial inversion symmetry, making them potential candidates for valley polarization. These materials’ properties, including valley polarization, have been investigated within single-particle density functional theory\cite{bc6n-1,hBN-valley-1}. However, a proper many-particle description of the excitons, that is necessary to capture the valley selective optical excitation properties is still lacking in the existing literature.

In this paper, we study the excitonic properties as well as capability of showing valley polarization of 2D BC$_6$N and B$_3$C$_2$N$_3$, using many-particle theory based on GW approximation and Bethe-Salpeter-equation (BSE). Monolayer BC$_6$N and B$_3$C$_2$N$_3$ exhibit strong bright exciton binding energies of 0.74 eV and 1.31 eV, respectively. Additionally, we find BC$_6$N and B$_3$C$_2$N$_3$ systems show a singlet-triplet splitting of 40 meV and 240 meV, respectively with higher triplet stability. Furthermore, the envelope function calculations along with the oscillator strength of valley selective optical transitions in presence of left- and and right-handed circular polarized lights show their perfect valley polarization behavior. Their strongly bound excitons within the moderate optical gap and ability to show valley polarization make them potential materials in photo-responsive information storage devices.

\section{Computational details}

\noindent We first calculate the ground state of these 2D borocarbonitrides using density functional theory (DFT), which is implemented in the Quantum Espresso code\cite{giannozzi2009quantum}. Here we consider the Perdew-Burke-Ernzerhof (PBE) exchange correlation function in the DFT calculations\cite{perdew1996generalized}. An energy cut-off of 100 Ry and a 12$\times$12$\times$1 Monkhorst-Pack \textbf{k}-point grid is used. The structure is optimized until the self-consistent energy difference between two successive electronic relaxation step is smaller than $10^{-10}$ eV and the forces on each atom become smaller than 0.01 eV/\AA. The vacuum layer is considered 20 \AA~ along the non-periodic direction to avoid any interaction between adjacent unit cells. Further, the quasiparticle band structure calculations are performed within the GW approximation using and BerkeleyGW package\cite{deslippe2012berkeleygw}. Here, G and W stand for Green's function and screened Coulomb potential, respectively, in the expression of Dyson's equation. For the GW calculations, the plane-wave energy cut-off is set at 100 Ry based on norm-conserving pseudopotentials that are taken from the Pseudo-Dojo database\cite{van2018pseudodojo}. We consider a 12$\times$12$\times$1 coarse \textbf{k}-grid, 912 bands and 12 Ry in dielectric cut-off for the converged GW calculations. A slab truncated Coulomb potential method is used to accelerate the convergence\cite{ismail2006truncation}. The BSE is solved for their optical absorption spectra using a 72$\times$72$\times$1 fine \textbf{k}-grid. The polarization of the incident light is considered to be parallel to the 2D plane of these systems. We consider zero center-of-mass momentum of the excitons in our BSE computation. We perform necessary convergence tests to obtain the parameters for both GW and BSE calculations.

\section{Results and Discussions}
\subsection{Berry curvature}

\begin{figure}[h]
\includegraphics[scale=0.43]{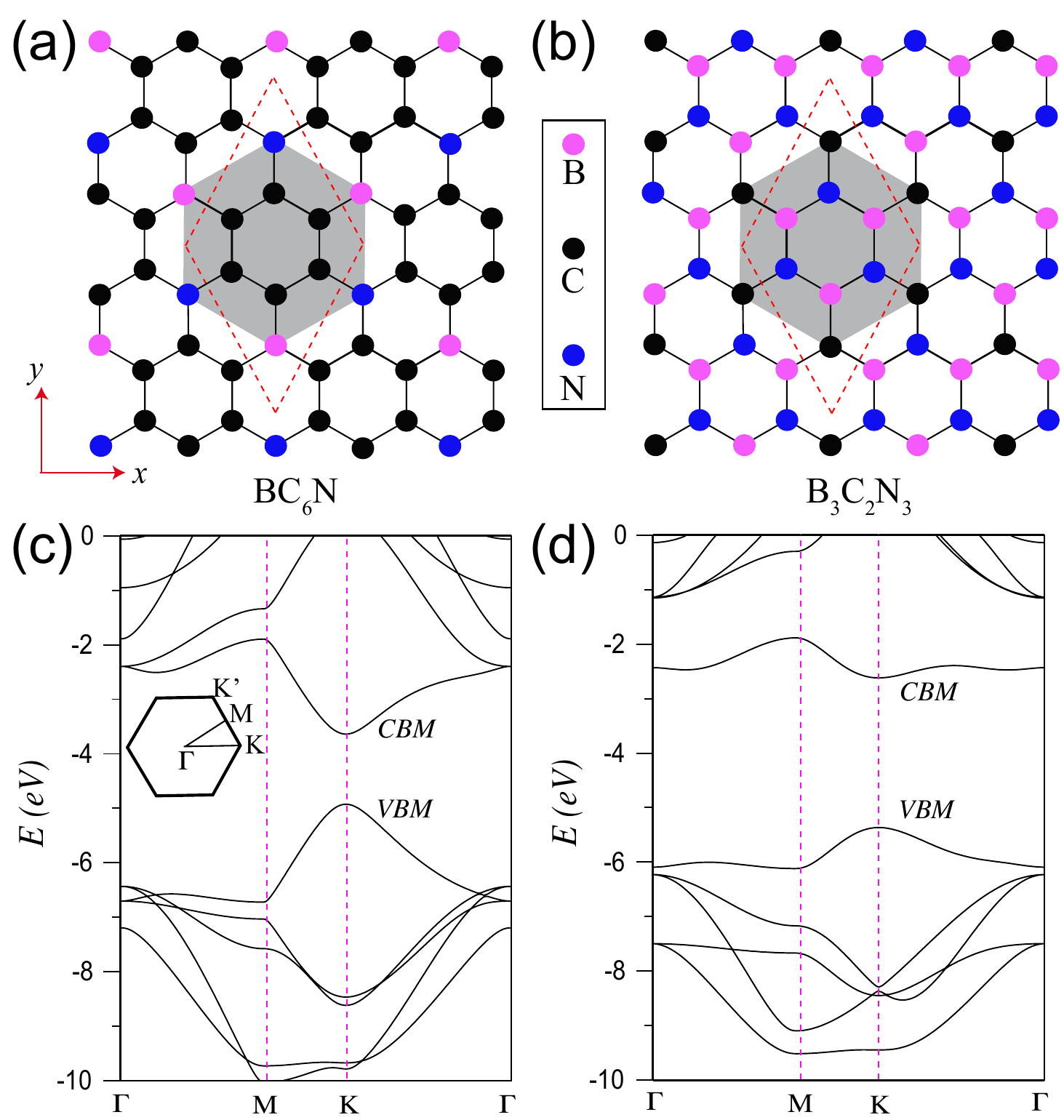}
\caption{\label{fig-1} The 2D honeycomb lattices of monolayer (a) BC$_6$N and (b) B$_3$C$_2$N$_3$. The immediate neighbors of carbon and alternate boron-nitrogen hexagons in BC$_6$N and B$_3$C$_2$N$_3$, respectively, are depicted by shaded hexagons. The dotted lines represent the rhombus unit cell of respective structures. The electronic band structure of (c)
BC$_6$N and  (d) B$_3$C$_2$N$_3$, as obtained using PBE exchange-correlation functional. The inset in (c) depicts the hexagonal first BZ with the high-symmetric points. The vertical dashed lines indicate the location of the high-symmetry points in the reciprocal space. The valence band maximum (VBM) and the conduction band minimum (CBM) are marked in the band structures.}
\end{figure}

\noindent We optimize the lattice structures of BC$_6$N and B$_3$C$_2$N$_3$ using DFT and depict the same in Fig.\ref{fig-1}a \& b. The planarity of these structures remain unaltered after the structural optimization. Both these structures are hexagonal with lattice constants 4.97 \AA~ and 5.02 \AA, respectively\cite{bc6n-1,karmakar2023g}. In the case of BC$_6$N carbon hexagon is surrounded by a hexagon with alternative boron and nitrogen atoms (grey colored shaded hexagon in Fig.\ref{fig-1}a). The structure of B$_3$C$_2$N$_3$ is just complementary to that of BC$_6$N. Here, boron nitride hexagon, made of alternate boron and nitrogen atoms is surrounded by a carbon hexagon (grey colored shaded hexagon in the Fig.\ref{fig-1}b).

Theoretical studies have previously demonstrated the high mechanical and dynamical stability of both these borocarbonitride systems\cite{bc6n-1,karmakar2023g}. These findings have been supported by experimental realizations of such systems using thermal catalytic chemical vapour deposition (CVD) techniques\cite{ci2010atomic}. In CVD, the heating temperature and duration can be controlled to adjust the composition of the material, along with the desired amounts of boron and nitrogen from respective sources. Further, Matsui \textit{et al.}, successfully synthesized 
boron-nitride doped nanographene with a similar atomic configuration to BC$_6$N\cite{matsui2018one}. This is achieved through a two-step process involving intra- and inter-molecular bora-Friedel-Crafts-type reactions. The success of these experimental syntheses further confirms the feasibility of realizing such borocarbonitride materials.

\begin{figure}[h]
\includegraphics[scale=0.45]{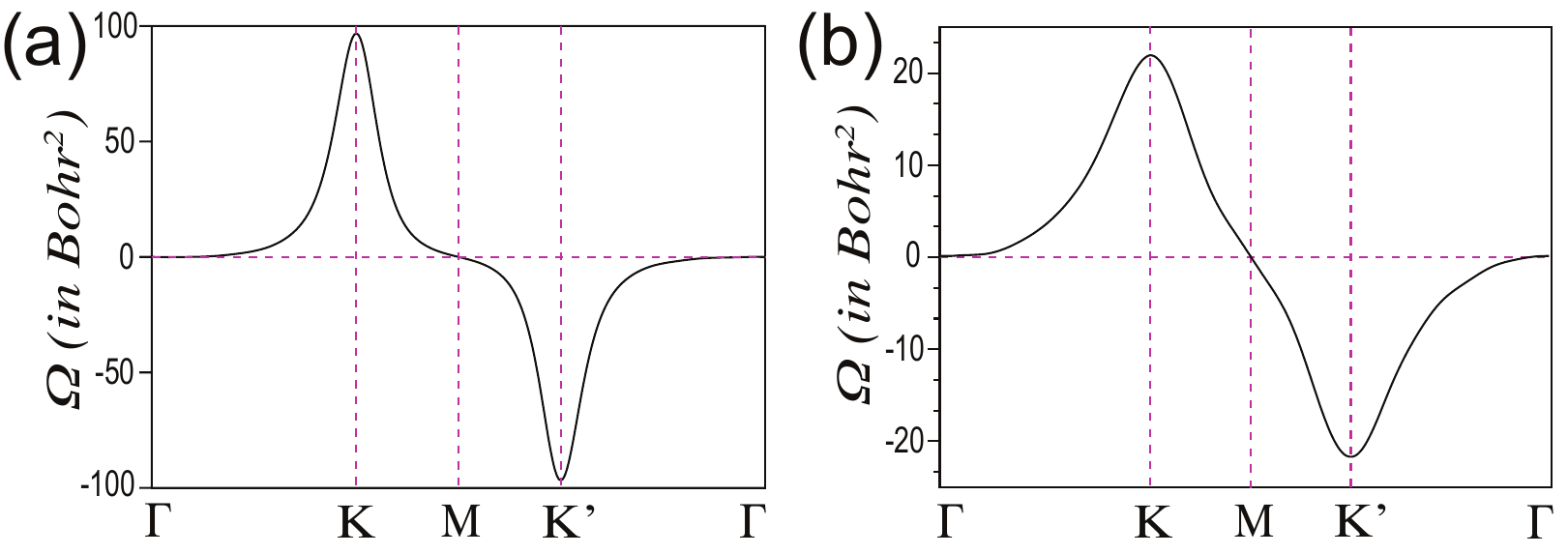}
\caption{\label{fig-2} The Berry curvatures of the top valence bands of (a) BC$_6$N and (b) B$_3$C$_2$N$_3$ along the high-symmetric points (dashed vertical lines) of the BZ.}
\end{figure}

We calculate their electronic band structures using PBE based density functional theory and present them in Fig.\ref{fig-1}c \& d, respectively. Both BC$_6$N and B$_3$C$_2$N$_3$ systems show a direct band gap of 1.28 eV and 2.74 eV, respectively at \textbf{K} 
high-symmetric point. Their band gaps within HSE06 functional\cite{heyd2003hybrid} consideration are reported to be 1.83 eV and 3.69 eV,
respectively\cite{bc6n-1,karmakar2023g}. Therefore, both structures are capable of showing good visible-to-ultraviolet absorption. Furthermore, the structures of both BC$_6$N and B$_3$C$_2$N$_3$ as depicted in Fig.\ref{fig-1}a \& b, clearly show the lack of spatial inversion symmetry that may lead to circular dichroism in these systems. We explore this property by calculating the Berry curvature corresponding to their valence band maxima within the DFT-based tight binding Hamiltonian and Wannier90 code as follows\cite{bc6n-1,pizzi2020wannier90}:

\begin{equation}
\Omega_z(\textbf{k}) = -\sum_{n\neq m}\frac{2 Im<\Psi_{n\textbf{k}}|v_x|\Psi_{m\textbf{k}}><\Psi_{m\textbf{k}}|v_y|\Psi_{n\textbf{k}}>}{(E_n - E_m)^2}
\end{equation}

\noindent where the summation is over all the valence bands, $v_{x(y)}$ is the velocity operator along the $x(y)$ direction and $E_{n(m)}$ are the $n(m)-th$ DFT eigenstate energy with $\Psi_{n(m)\textbf{k}}$ valence eigenstate. The calculated Berry curvatures of both the structures are shown in Fig.\ref{fig-2}a \& b, respectively. Owing to their broken spatial inversion symmetry, both BC$_6$N and B$_3$C$_2$N$_3$ show opposite Berry curvatures at two valleys, \textit{i.e.}, at \textbf{K} and \textbf{K}$'$ high-symmetric points in the first BZ, with peak values of 96 Bohr$^2$ and 22 Bohr$^2$, respectively. These large values of opposite Berry curvatures are proportional to the effective magnetic field in these two valleys. Under the influence of such effective magnetic field, the electrons from the opposite valleys
couple with opposite chirality of circular polarized lights, resulting in valley selective electronic excitations. This phenomenon is known as valley polarization. The Berry curvature calculations are perfectly matched with the previous results, where they have used single-particle based DFT\cite{bc6n-1,hBN-valley-1}. However, as seen in the case of the monolayer 2D h-BN system, the quasiparticle character of the excitons and associated intervalley scattering can significantly suppress the valley polarization ability\cite{hbn-prl-intervalley}.
To capture the reduced screened Coulomb interactions along with the intervalley scattering processes, we further calculate their valley excitonic quasiparticle properties within the many-body perturbation theory based on GW plus BSE formalism.

\subsection{Quasiparticle band structures}

\begin{figure}[h]
\includegraphics[scale=0.4]{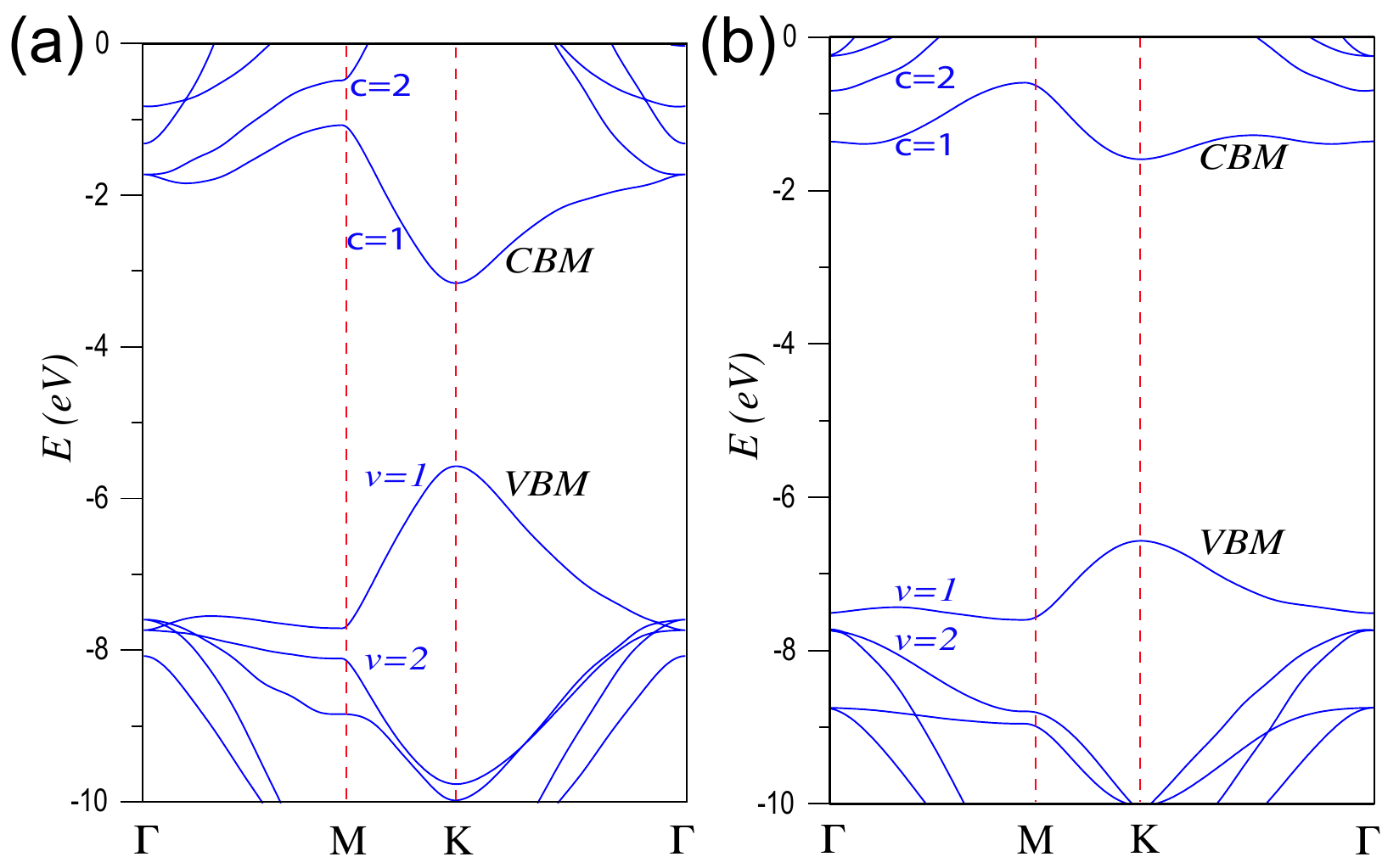}
\caption{\label{fig-3} The GW band structures of (a) BC$_6$N and (b) B$_3$C$_2$N$_3$. The vertical dashed lines depict the positions of the high-symmetric points in the BZ. The valence ($\nu$) and conduction bands ($c$) are marked with numerical values, starting from the frontier bands.}
\end{figure}

\noindent We calculate the quasiparticle (QP) band dispersion of both the systems within GW approximation as follows\cite{GW-method-1}:
\begin{equation}
 [-\frac{1}{2}\nabla^2 + V_{ext} + V_H + \Sigma(E^{QP}_{n\textbf{k}})]\Psi^{QP}_{n\textbf{k}}  = E^{QP}_{n\textbf{k}}\Psi^{QP}_{n\textbf{k}} 
\end{equation}
where $E^{QP}_{n\textbf{k}}$ and $\Psi^{QP}_{n\textbf{k}}$ are the quasiparticle energies and wave functions, respectively. The first term is the kinetic energy operator of electrons considering $\hbar$ = $m$ (mass of electron) = 1. The second and third terms are the potential energy due to the ion-electron interactions and the Hartree energy, respectively. The fourth term is the self-energy operator of electrons within the GW approximations. In the Fig.\ref{fig-3}a \& b, we depict the quasiparticle band structures of both BC$_6$N and B$_3$C$_2$N$_3$, respectively. The electronic self-energy corrections have a direct impact on the dramatically enhanced band gaps. The BC$_6$N and B$_3$C$_2$N$_3$ systems show a GW gap (within the G$_0$W$_0$ level) of 2.41 eV and 4.98 eV at the \textbf{K} high-symmetric point, respectively. The self-consistency of our results are further confirmed by one additional self-consistent update for G, \textit{i.e.}, G$_1$W$_0$, which results into the same band gaps for the both structures.

\subsection{Excitonic absorption spectra and circular dichroism}

\noindent We further explore their excitonic properties from the two-particle BSE formalism as follows\cite{BSE-1}:

\begin{equation}
(E^{QP}_{c\textbf{k}} - E^{QP}_{\nu\textbf{k}})A^S_{\nu c \textbf{k}} + \sum_{\nu'c'\textbf{k}'} <\nu c \textbf{k}|K^{e-h}|\nu'c'\textbf{k}'> = \Omega^S A^S_{\nu c \textbf{k}}
\end{equation}

\noindent where $\nu$(c) represents the valence (conduction) band-index, $A^S_{\nu c\textbf{k}}$ is the eigenvector of exciton, $\Omega^S$ is the electron-hole (e-h) excitonic energy and $K^{e-h}$ is the e-h interaction kernel. The term $E^{QP}_{c\textbf{k}}$ and $E^{QP}_{\nu\textbf{k}}$ represent the quasiparticle energies of the conduction and valence states, respectively. After solving the above BSE, the excitonic absorption spectra is obtained from the imaginary part of dielectric function:

\begin{equation}
\epsilon_2 (\omega) = \frac{16\pi^2e^2}{\omega^2}\sum_S|\textbf{P}.<0|\textbf{v}|S>|^2\delta(\omega -\Omega^S)
\end{equation}

\noindent where $|S>$ is the excitonic wave function, $|0>$ is the Fock space within the DFT level, $\omega$ is the incident photon energy, $\textbf{P}$ is the polarization vector, $\textbf{v}$ is the velocity operator, and $e$ is the electronic charge. 

\begin{figure}[h]
\includegraphics[scale=0.4]{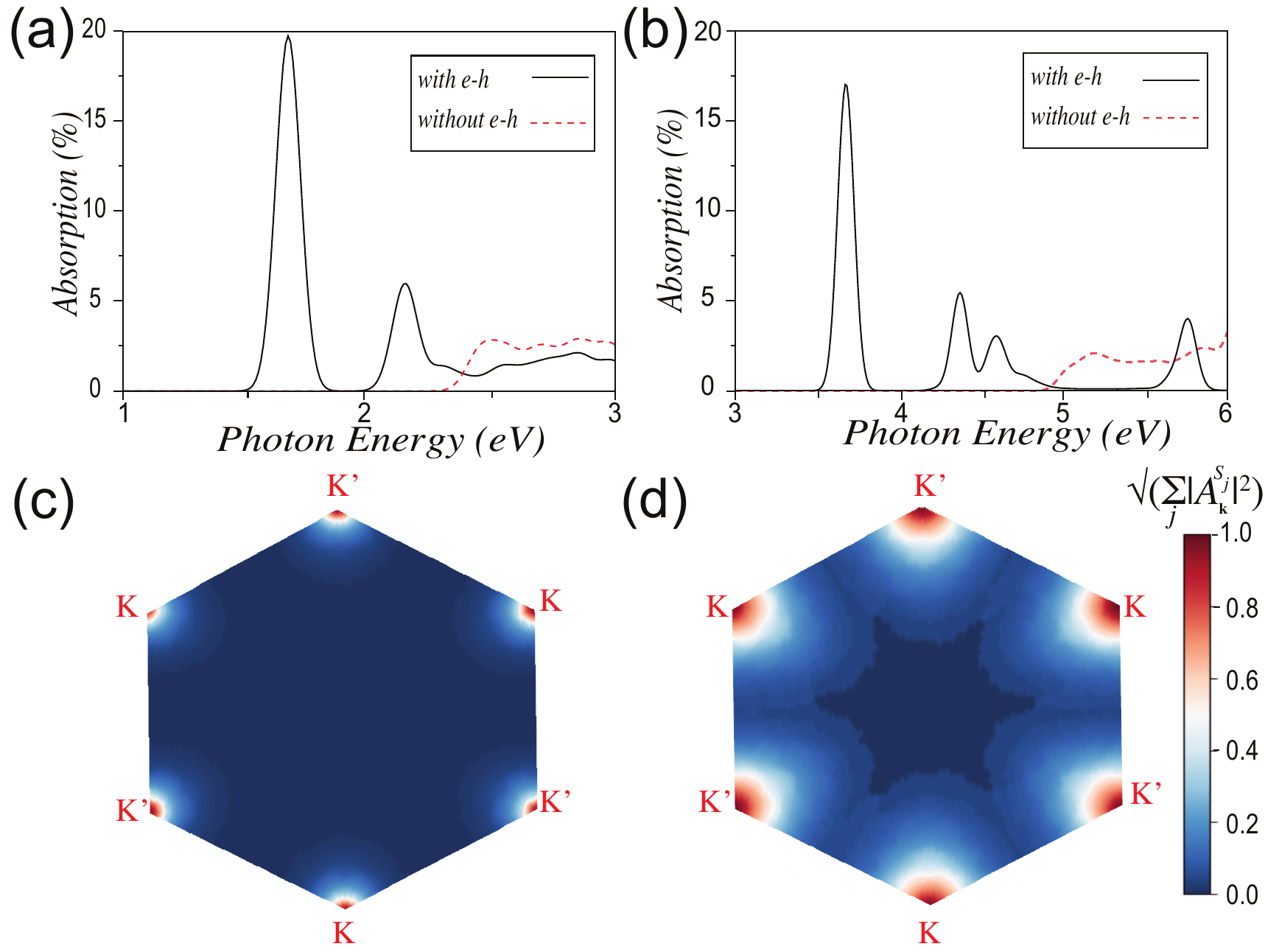}
\caption{\label{fig-4} The optical absorption spectra of (a) BC$_6$N and (b) B$_3$C$_2$N$_3$ with (soild lines) and without (dashed lines) quasi electron and quasi-hole interactions. using a constant broadening of 50 meV. Here, the excitons with less than 5\% of the oscillator strength of the brightest exciton are defined as dark excitons. The contour plots of square root of the \textbf{k}-resolved envelope function of first bright exciton of (c) BC$_6$N and (d) B$_3$C$_2$N$_3$, over hexagonal first BZ. The color bar shows the normalized values of the envelope function.}
\end{figure} 

The optical absorption spectra in the presence of linear polarized light of both the structures are presented in Fig.\ref{fig-4}a \& b. The convergence of these calculations are achieved by varying the number of valence and conduction bands. As can be seen, the first bright singlet excitonic peak in the optical absorption spectrum with quasi-electron and quasi-hole interactions (black solid line) can be seen at 1.67 eV and 3.67 eV in the cases of BC$_6$N and B$_3$C$_2$N$_3$, respectively. The absorption spectra without the electron-hole interactions (red dotted line) begin at 2.41 eV for BC$_6$N and at 4.98 eV for B$_3$C$_2$N$_3$. The difference between the energies with and without quasi-hole and quasi-electron interactions gives the excitonic binding energy. The binding energy of the first bright exciton of BC$_6$N and B$_3$C$_2$N$_3$ are 0.74 eV and 1.31 eV, respectively. The binding energies of both the structures are comparable to that of monolayer MoS$_2$ (0.96 eV)\cite{qiu2013optical}, blue phosphorene (0.84 eV)\cite{zhou2021unusual}, and black phosphorene(0.78 eV)\cite{tran2015quasiparticle}. Furthermore, the eigenvalues of BSE show double degeneracy of the first bright exciton for both the systems. Such degeneracy of these excitons can be broken by irradiating different chirality of circularly polarized light, as it has been ovserved in case of monolayer MoS$_2$\cite{qiu2013optical}. This can lead to the manifestation of valley polarization.

The valley polarization of a system can be identified by the oscillator strength of an exciton in presence of different chirality of circularly polarized light. However, the intervalley coupling plays a significant role in determining whether a system can exhibit valley polarization, as observed in the case of monolayer h-BN\cite{hbn-prl-intervalley}. In monolayer h-BN, the presence of significant intervalley coupling results in the suppression of valley polarization. Therefore, here we calculate the \textbf{k}-resolved envelope function: $A^{S_j}_{\nu c \textbf{k}}$ ($j$ is the degeneracy of the exciton) of the first bright exciton of both the systems in presence of linear polarized light. This envelope function is directly related to the scaled oscillator strength of an exciton as follows\cite{hbn-prl-intervalley}: $I = \sum_{j}|A^{S_j}_{\nu c \textbf{k}}\textbf{P}.<\nu\textbf{k}|\textbf{v}|c\textbf{k}>|^2$. The term in the oscillator strength, $\textbf{P}.<\nu\textbf{k}|\textbf{v}|c\textbf{k}>$ decides the transition possibility at a given $\textbf{k}$. In Fig.\ref{fig-4}c \& d we plot the \textbf{k}-resolved envelope functions of BC$_6$N and B$_3$C$_2$N$_3$ in their first BZ, respectively. As can be seen, the first bright exciton of BC$_6$N is localized around the \textbf{K} and \textbf{K}$'$ high-symmetric points of the BZ. There is no inter valley coupling between the two valleys. On the other hand, the first bright exciton of B$_3$C$_2$N$_3$ shows small inter valley coupling of 5.6\% in between the \textbf{K} and \textbf{K}$'$ valleys.

\begin{figure}[h]
\includegraphics[scale=0.45]{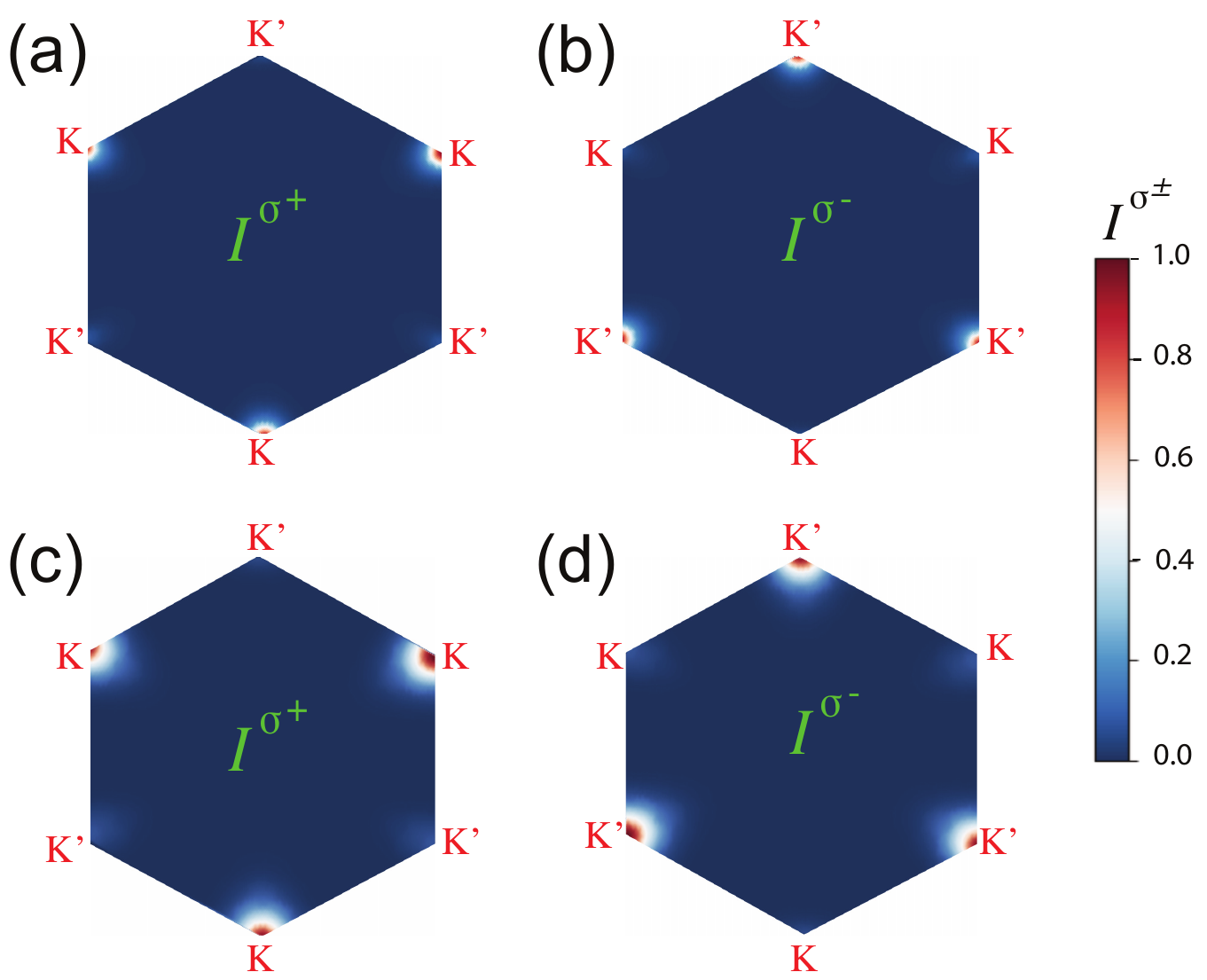}
\caption{\label{fig-5}The oscillator strengths of the first bright exciton of BC$_6$N in presence of (a) left- and (b) right-handed circular polarized lights. The same oscillator strength plots in the case of B$_3$C$_2$N$_3$ in (c) and (d). The color bar depicts the normalized values of the oscillator strength. }
\end{figure}

As we have mentioned before, the first bright excitons of both BC$_6$N and B$_3$C$_2$N$_3$ are arising from two degenerate excitonic states. One of them is localized around the \textbf{K} valley
and the other one is localized around the \textbf{K}$'$ valley. These degeneracy can be broken by the different handed circular polarized light. Further, to check their valley polarization capabilities, the scaled oscillator strength $I$ can be rewritten for the left $I^{\sigma^{+}}$
and right $I^{\sigma^{-}}$ handed circular polarized light as follows\cite{hbn-prl-intervalley}:

\begin{equation}
I^{\sigma^{\pm}} = \sum_{j}|A^{S_j}_{\nu c\textbf{k}}\textbf{P}.<\nu\textbf{k}|(v_x\hat{x} \pm iv_y\hat{y})|c\textbf{k}>|^2
\end{equation}

\noindent where $v_x$ and $v_y$ are the components of the velocity operator along the $x$- and $y$-directions, respectively and $\hat{x}$ ($\hat{y}$) is the unit vector of the velocity along the $x$ ($y$)-axis. Using the above equation we plot the oscillator strengths of the first bright excitons of both BC$_6$N and B$_3$C$_2$N$_3$ in presence of left- and right-handed circularly polarized lights in the Fig.\ref{fig-5}.
We can see that the excitons at \textbf{K}$'$ and \textbf{K} valleys of BC$_6$N and B$_3$C$_2$N$_3$ are formed selectively by the left- and right-handed circularly polarized lights, respectively. Therefore, in the presence of an in-plane electric field, the electrons at \textbf{K}$'$ and \textbf{K} valleys will experience opposite Lorentz force and they will prefer to move in opposite directions. These observations establish both the BC$_6$N and B$_3$C$_2$N$_3$ systems as efficient valley polarizer that are capable of exhibiting circular dichroism behavior and can be exploited for valley-Hall device fabrication.

\subsection{Triplet exciton formation}

\begin{figure}[h]
\includegraphics[scale=0.45]{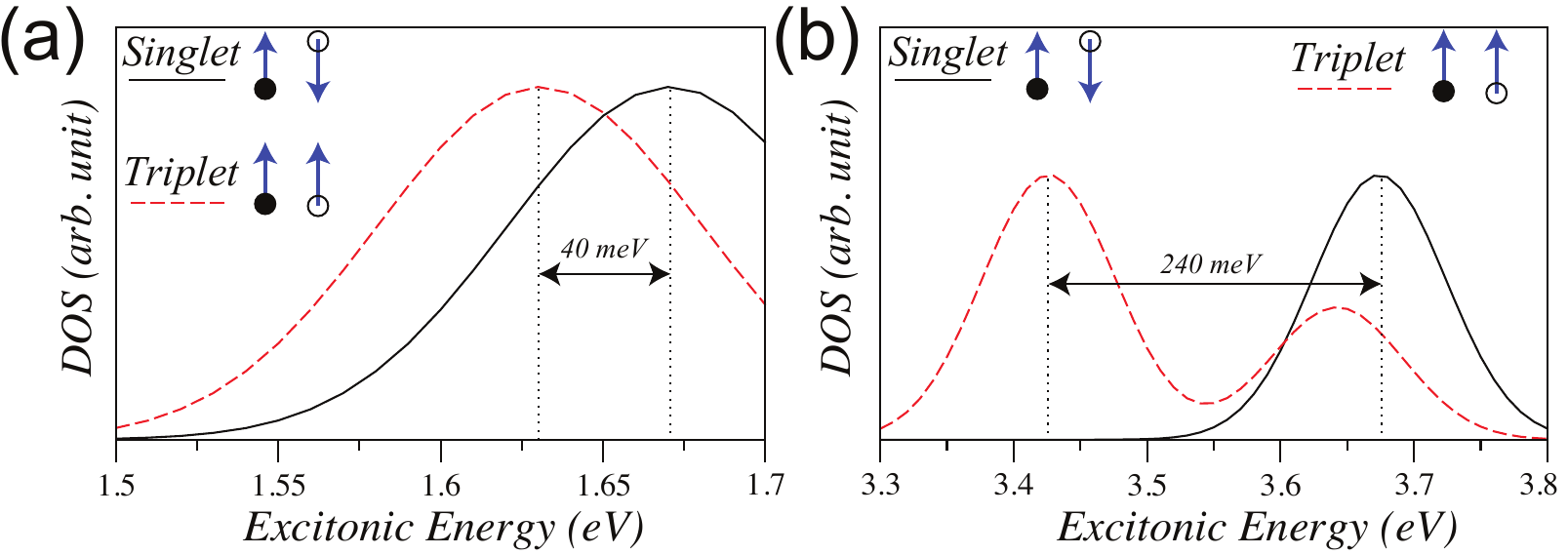}
\caption{\label{fig-6} The DOS of singlet (solid lines) and triplet (dashed lines) excitons of (a) BC$_6$N and (b) B$_3$C$_2$N$_3$. The vertical dotted lines indicate the location of the first singlet and first triplet exciton peaks.}
\end{figure}

\noindent We further explore the features of triplet excitons. Note that, the triplet excitons that are formed by parallel alignment of electron and hole spins, are always dark due to the forbidden optical selection rule\cite{BSE-1}. We plot the normalized density of state (DOS) of both
singlet and triplet excitons for both the BC$_6$N and B$_3$C$_2$N$_3$ systems, in the Fig.\ref{fig-6}a \& b, respectively. In the case of BC$_6$N, the positions of first bright singlet and first triplet excitons appear at different energy values as shown by the two vertical dotted lines in the Fig.\ref{fig-6}a with a singlet-triplet splitting of 40 meV. Similarly, in the case of B$_3$C$_2$N$_3$, the DOS plot in Fig.\ref{fig-6}b shows a large singlet-triplet splitting of 240 meV. Such high singlet-triplet splitting ensures higher stability of the triplet excitons at finite temperatures, leading to an enhanced quantum efficiency for the optical device applications\cite{triplet-stability-1,triplet-stability-2}. The triplet exciton always originates from the exchange energy between the quasi-electron and quasi-hole\cite{BSE-1}. 

To gain further insight of the origins of these singlet and triplet excitons we calculate the band-to-band transitions ($T_{tot}$). We mark the valence and conduction bands in Fig.\ref{fig-3}a \& b, by $\nu$ and $c$, respectively with ascending order of numerical indices, starting from the frontier bands. The total excitation probability is calculated by the formula: $T_{tot} = \sum_{\textbf{k}} |A_{\nu c \textbf{k}}|^2$, where $A_{\nu c \textbf{k}}$ is the amplitude of an exciton at a specific momentum $\textbf{k}$. We find that, the formation of
singlet excitons in both the systems occurs majorly due to the optical transition probability of 99.99\% from top of the valence band ($\nu$ = 1) to the bottom of the conduction band 
($c$ = 1). The contributions towards the triplet exciton formation from these frontier bands are 99.86\% and 97.78\% in case of BC$_6$N and B$_3$C$_2$N$_3$, respectively. However, in case of B$_3$C$_2$N$_3$, a contribution of 2.22 \% appear from the optical transitions between non-frontier bands. This difference gives rise to a larger singlet-triplet splitting in B$_3$C$_2$N$_3$ as compared to BC$_6$N.

\section{Conclusions}

\noindent In conclusion, we investigate the excitonic and valley properties of the 2D noncentrosymmetric BC$_6$N and B$_3$C$_2$N$_3$ through first-principles calculation based many-body perturbation theory. We find 1.67 eV and 3.67 eV optical gaps with 0.74 eV and 1.31 eV excitonic binding energies of BC$_6$N and B$_3$C$_2$N$_3$, respectively. These optical gaps indicate efficient optical absorption in the visible to ultraviolet region. Furthermore, these systems demonstrate the capability to stabilize the triplet exciton over the singlet exciton by 40 meV and 240 meV, respectively, which has the potential to enhance the exciton lifetimes and the quantum efficiency of optical absorption. Due to their broken spatial inversion symmetry, these systems exhibit opposite Berry curvatures at time-reversal pair valleys \textit{i.e.}, \textbf{K} and \textbf{K}$'$ high-symmetric points. The 2D BC$_6$N show zero intervalley coupling for its first bright exciton, whereas the B$_3$C$_2$N$_3$ system shows a negligible intervalley coupling between the valleys. However, the corresponding oscillator strengths clearly indicate that the qusasiparticles in opposite valleys can be excited selectively by controlling the chirality of the circular polarized light for both the systems. This property can be leveraged for the realization of excitonic qubits through circular dichroism valley-Hall devices.

\section{Acknowledgment}
SA and SD thank IISER Tirupati for Intramural Funding and Science and Engineering Research Board, Dept. of Science and Technology, Govt. of India for research grant (CRG/2021/001731). The authors acknowledge National Supercomputing Mission (NSM) for providing computing resources of ‘PARAM Brahma’ at IISER Pune, which is implemented by C-DAC and supported by the Ministry of Electronics and Information Technology (MeitY) and DST, Govt. of India.

\providecommand{\latin}[1]{#1}
\makeatletter
\providecommand{\doi}
  {\begingroup\let\do\@makeother\dospecials
  \catcode`\{=1 \catcode`\}=2 \doi@aux}
\providecommand{\doi@aux}[1]{\endgroup\texttt{#1}}
\makeatother
\providecommand*\mcitethebibliography{\thebibliography}
\csname @ifundefined\endcsname{endmcitethebibliography}
  {\let\endmcitethebibliography\endthebibliography}{}

\end{document}